\documentstyle[prl,aps,epsf]{revtex}
\draft
\begin{document}
\twocolumn[\hsize\textwidth\columnwidth\hsize\csname @twocolumnfalse\endcsname

\title{Phase-locking of driven vortex lattices with transverse ac-force and
periodic pinning}
\author{Charles~Reichhardt$^1$, Alejandro~B.~Kolton$^2$, Daniel~Dom\'{\i}nguez$^2$, and Niels~Gr{\o}nbech-Jensen$^{3,4}$}
\address{$^1$Applied Physics Division and Center for Nonlinear Studies, Los Alamos National Laboratory, New Mexico 87545\\
$^2$Centro Atomico Bariloche, 8400 S.~C.~de~Bariloche, Rio Negro, Argentina\\
$^3$Department of Applied Science, University of California, Davis, California 95616\\
$^4$NERSC, Lawrence Berkeley National Laboratory, Berkeley, California 94720}

\date{\today}
\maketitle
\begin{abstract}
For a vortex lattice moving in a periodic array we show analytically and 
numerically 
that a new type of phase locking occurs in the presence of  
a longitudinal dc driving force and a transverse ac driving force. 
This phase locking is distinct from the Shapiro 
step phase locking found with longitudinal ac drives. 
We show that an increase in critical current and a fundamental phase locked  
step width scale with the square of the driving ac amplitude. 
Our results should carry over to other systems such as vortex motion
in Josephson-junction arrays. 
\end{abstract}
\vspace{-0.1in}
\vspace{-0.3in}
\vskip2pc]
\narrowtext
\section{Introduction}
When an external ac drive is applied to a dc driven system
interacting with a
periodic potential, resonance between the external ac frequency
and the frequency of motion in the periodic potential can give  
rise to phase locking. This 
phenomenon is found in a wide variety of
nonlinear systems in condensed matter physics. 
A particularly well known example is  the
ac/dc-driven single small
Josephson-junction \cite{Shapiro,Barone} and 
Josephson-junction arrays \cite{ShapiroN},  
where Shapiro steps are observed in
the current-voltage characteristics. 
Shapiro type phase locking has recently also been 
studied in experiments \cite{VanLook} 
and theory \cite{Shapiro2} 
for vortex motion in two-dimensional superconductors
with periodic arrays of pinning sites, as well as in
superconductors where the vortices
are driven over a periodic potential generated by 
thickness modulations \cite{Daldini}. 
Moreover, driven systems with many degrees of freedom in the
presence of quenched disorder can exhibit phase-locking
when there is a dynamically induced periodicity.
Examples of this are charge density waves \cite{CDW} 
and vortex lattices in superconductors with random pinning \cite{Fiory,Kolton}.
In the case of vortices in superconductors
with periodic pining \cite{VanLook,Shapiro2,Daldini},  
a  vortex can be viewed as an overdamped particle  
moving along a tilted washboard potential
where the tilt is produced by the dc force from 
an applied current as well as by the  superimposed 
ac currents in the same direction. 

An important difference
of vortex arrays from charge density wave systems 
and single degree of freedom systems (like a small Josephson
junction) is that the displacement field acting on vortices is
a two-dimensional vector. This means that displacements can be
induced in two different directions, and therefore a new kind
 of phase locking is possible when an ac force
is applied {\it transverse} to the direction of the dc force.

In this work we show analytically and numerically 
how phase locking phenomena may occur
when vortices are moving in a periodic potential 
and the ac force is applied transverse to the direction of the
longitudinally applied dc force. 
The type of phase-locking observed in this case is qualitatively different
from the Shapiro type phase-locking observed when dc and ac forces are
in parallel. We first show analytically the possibility of the phase locked 
states for certain commensurate vortex configurations and predict that the
critical current {\it increases} quadratically
with the ac amplitude and that the width of some of the phase-locked
steps scale as the square of the ac amplitude. 
We also find scaling of the critical current as well as the first phase 
locked region as a function of the ac frequency and the pinning geometry. 
The predictions of the perturbation analysis are confirmed with numerical
simulations. We analyze the validity
of the perturbation approach and demonstrate how deviations occur.
Our results suggest that more
pronounced transverse phase-locking may be observed at 
lower commensurate magnetic fields, such as $B=1.5B_\phi$ or $B=1.25B_\phi$.
The general model and perturbation results are easily generalizable to
other systems which exhibit Shapiro step-like phase locking such as 
Josephson-junction arrays and Frenkel-Kontorova type models.

Our studies are directly relevant for several contemporary efforts
since vortex matter   interacting with nanostructured pinning arrays
of holes \cite{Metlushko,Moshchalkov,Harada} 
and dots \cite{Schuller,Hoffman} has been attracting
increasing attention due to the easily tunable pinning properties.  
Pronounced commensuration effects are observed in these systems 
when the density of vortices matches to integer or fractional multiples of the
density of pinning sites. In addition to square pinning arrays,
recent experiments have been conducted on 
rectangular pinning arrays \cite{Hoffman}. 
Simulations have shown several interesting dynamical phases of dc driven
vortices in periodic pinning systems
\cite{DrivenShort,Marconi}.
Imaging \cite{Harada} and transport experiments \cite{Moshchalkov,VanLook} 
along with simulations \cite{DrivenShort} have found that
the vortex motion above the first matching field can occur by the flow
of interstitial vortices between the pinning sites. These vortices still 
experience a periodic potential created by the vortices located at the
pinning sites. Recently, phase locking was observed for dc and ac driven
vortices
interacting with periodic pinning at $B=2B_{\phi}$, where $B_{\phi}$
is the field at which the number of vortices equals the number of pinning
sites \cite{VanLook}, where the dc and ac drive where in the same direction.
The values of the voltage response in Ref.~\cite{VanLook} strongly
suggest that is is the interstitial vortices that are mobile while the
vortices on pinning sites remain immobile.

\section{The Model}

We will adopt a simple overdamped particle model for describing vortices.
The overdamped normalized equation of motion for vortex $i$ is 
\begin{eqnarray}
\dot{{\bf r}}_i & = & -\nabla_iU(\{{\bf r}_j\}) + {\bf f}_d \; ,
\label{eq:Eq_1}
\end{eqnarray}
where ${\bf r}_i = (x_i,y_i)$ is the coordinate of vortex $i$ normalized
to the effective 2D penetration depth, $\Lambda=2\lambda^2/d$, $\lambda$ being
the penetration depth and $d\ll\lambda$ is the sample thickness. The
energy surface, $U$, is normalized to $A_v=\Phi_0^2/8\pi\Lambda$, where
$\Phi_0$ is the flux quantum, and the external driving force,
${\bf f}_d={\bf i}\times\hat{\bf z}$, where ${\bf i}$ is the normalized
bias current and
$\hat{\bf z}$ is a normalized unit vector. Time is normalized to
$\tau=\alpha\Lambda/A_v$, where $\alpha$ is the Bardeen-Stephen friction
coefficient.
The energy surface, $U=U_{vv}(\{{\bf r}_j\})+U_{vp}(\{{\bf r}_j\})$,
is broken into vortex-vortex interactions, $U_{vv}(\{{\bf r}_j\})$, and
vortex-pinning interactions, $U_{vp}(\{{\bf r}_j\})$.
Vortex-vortex interactions are modeled as pair interactions,
\begin{eqnarray}
U_{vv}(\{{\bf r}_j\}) & = &
\sum_{i=1}^{N_v}\sum_{j\neq i}^{N_v}u_{vv}(r_{ij}) \; ,
\label{eq:Eq_2}
\end{eqnarray}
where $r_{ij}=|{\bf r}_j-{\bf r}_i|$, and vortex-pinning interactions
are modeled by,
\begin{eqnarray}
U_{vp}(\{{\bf r}_j\}) & = &
\sum_{i=1}^{N_v}\sum_{j=1}^{N_p}u_{vp}(r_{ij}) \; ,
\label{eq:Eq_3}
\end{eqnarray}
where $N_v$ and $N_p$ are the number of simulated vortices and pinning
centers, respectively.

We are performing simulations with periodic
boundary conditions (periodicity $L_x$, $L_y$),
and with a logarithmic interaction potential ($u_{vv}(r_{ij})=-\ln{r_{ij}}$)
valid when $\Lambda$ is of the order of the sample size.

Combining the logarithmic interactions with the periodic boundary conditions,
result in \cite{Jensen2}
\begin{eqnarray}
&& u_{vv}(x_{ij},y_{ij}) \; = \; C\left(\frac{L_y}{L_x}\right)+\pi\frac{L_y}{L_x}\left(\frac{y_{ij}}{L_y}\right)^2 \label{eq:Eq_4} \\
& &- \frac{1}{2}
\ln\left[\prod_{n=-\infty}^\infty\frac{\cosh\left(2\pi\frac{L_y}{L_x}(\frac{y_{ij}}{L_y}+n)\right)-\cos\left(2\pi\frac{x_{ij}}{L_x}\right)}{\cosh\left(2\pi\frac{L_y}{L_x}n\right)}\right] \; , \nonumber
\end{eqnarray}
where $(x_{ij},y_{ij})$ are the cartesian coordinates,
$r_{ij}^2=x_{ij}^2+y_{ij}^2$, and $C(\cdot)$ is a function determined by
the energy normalization of the long range interactions.

We have previously demonstrated that phase-locking of the vortex motion
can exist if a certain periodicity of the global pinning potential is
present. This was shown for driving forces of the form,
${\bf f}_d=(\eta+\varepsilon\sin{\Omega t})\hat{{\bf x}}$, where
$\hat{{\bf x}}$ is a unit vector in the x-direction. Analogous to the
well-known Shapiro steps \cite{Shapiro} in the current-voltage characteristics
of an ac/dc-driven Josephson junction, it was demonstrated that nonzero
intervals of $\eta$ will result in phase-locking of vortex motion with
vortex velocities resonating with the external ac-field. We will in this
paper demonstrate that one can also expect phase-locking in the transversely
ac-driven case, i.e., for
\begin{eqnarray}
{\bf f}_d & = & \left\{\begin{array}{c}\eta \\ \varepsilon\sin{\Omega t} \; ,
\end{array}\right\}
\label{eq:Eq_5}
\end{eqnarray}
where $\eta$ and $\varepsilon$ are the magnitudes of dc and ac forces and
$\Omega$ is the frequency of the ac drive.
The relevant equations of motion are therefore,
\begin{eqnarray}
\dot{x}_i + \frac{\partial}{\partial x_i} U(\{{\bf r}_j\}) & = & \eta \label{eq:Eq_6}\\
\dot{y}_i + \frac{\partial}{\partial y_i} U(\{{\bf r}_j\}) & = & \varepsilon\sin{\Omega t}
\label{eq:Eq_7}
\end{eqnarray}

\section{Analytical Results}

We will consider a system with periodic pinning in a rectangular
periodic lattice with plaquette area $A=l_xl_y$, where $l_x$ and $l_y$ are
the dimensions of the rectangular plaquettes. We will assume that the
applied magnetic
field is larger than the first matching field, such that all pinning
centers are occupied by trapped vortices. We will in the following assume
that the trapped vortices remain trapped at the pinning sites and
immobile at all times. Thus the interstitial vortices move in a strictly
periodic lattice of immobile vortices with periodicity $(l_x,l_y)$.
We can then perform a perturbational analysis of the interstitial vortex
behavior through the following assumption:
any interstitial vortices moves such that the forces from other interstitial
vortices cancel by symmetry \cite{Shapiro2}.
Thus, analyzing the behavior of the interstitial (mobile)
vortices, we will adopt the potential function for a single interstitial
vortex moving in a lattice of pinned vortices (at positions
$(nl_x,(m+\frac{1}{2})l_y)$), as given in Eq.~(\ref{eq:Eq_4}).
Thus, the interstitial vortex position is subject to
a potential as in Eq.~(\ref{eq:Eq_4}) where the coordinate is the vortex
position, $(x_{i},y_{i})$, and with the lattice constants here being the
distances between pinning centers, $(l_x,l_y)$, instead of the computational
system dimensions, $(L_x,L_y)$.

\subsection{Critical dc-force without ac-drive}

The first result to evaluate for the purely dc-driven case is the ``critical
dc current", $\eta_c^{(0)}$, below which the interstitial vortex
is trapped by the pinned vortex lattice. As $\eta$ is increased from $\eta=0$,
the interstitial vortex will find an equilibrium point for which the
pinned vortex lattice provides a cancelling force to the dc bias. However,
for a given (critical) value of $\eta$, namely $\eta_c^{(0)}$, the pinned vortex
lattice can no longer provide enough force to resist motion of the interstitial,
which therefore begins to propagate in the x-direction. Assuming that the
pinned vortices do not move and that an interstitial vortex only interacts
with pinned vortices, this critical force can be found as the
maximum gradient value of Eq.~(\ref{eq:Eq_4}) for $y_{ij}=ml_y$.
The corresponding equation of motion is given by
Eq.~(\ref{eq:Eq_6}):
\begin{eqnarray}
\dot{x}_i + \frac{2\pi}{l_x}\sum_{n=-\infty}^\infty\frac{\sin\left(2\pi\frac{x_i}{l_x}\right)}{\cosh\left(2\pi\frac{l_y}{l_x}(\frac{1}{2}+n)\right)-\cos\left(2\pi\frac{x_i}{l_x}\right)} & = & \eta \; . \nonumber \\
\label{eq:Eq_8}
\end{eqnarray}
This equation can produce a critical current by requiring $\dot{x}_i=0$ and
optimizing the left hand side through the varying the position, $x_i$.
In general, this must be done numerically.
The result is shown in Fig.~(1), where the critical dc force (thick solid)
is shown together with the optimized value of the equilibrium value of
$x_i=x_*$. In the limit of
${\rm sech}(\pi l_y/l_x)\ll 1$, we can derive the approximate expression
of the critical current by noticing that only two terms ($n=-1,0$) are
required to describe the critical current, which is then always given
by $x_i=x_*=\frac{3}{4}l_x$:
\begin{eqnarray}
\eta_c^{(0)} & = & k{\rm sech}\left(\pi\frac{l_y}{l_x}\right) \; \; , \; \; {\rm for} \; \; \frac{\pi}{2}\gg\frac{l_x}{l_y}
\label{eq:Eq_9} \\
k & = & \frac{2\pi}{l_x} 
\label{eq:Eq_10}
\end{eqnarray}

The approximate expression, Eq.~(\ref{eq:Eq_9}), is shown in figure 1
as a thin solid curve,
validating that the interaction potential between interstitial and pinned
vortices can be simplified to the terms $n=-1,0$ for
$l_x{{<}\atop{\sim}}1.5l_y$. One can also see that in the extreme opposite
limit, $l_x\gg l_y$, the critical dc force, $\eta_c^{(0)}$, can be approximated
by $\pi/l_y$. This is shown as a dashed line in figure 1.

It is here important to remember that the analysis, which is based on
a magnetic field of ``matching field plus one flux quantum",
is actually valid for magnetic fields, $B_\phi<B\le2B_\phi$,
for which the vortex configuration results on forces between interstitial
vortices cancelling due to symmetry; such magnetic fields can be, e.g.,
$B=2B_\phi,\frac{3}{2}B_\phi,\frac{5}{4}B_\phi,\cdots$.

\subsection{Transverse ac-drive included}

Let us assume that ${\rm sech}\left(\pi\frac{l_y}{l_x}\right)\ll1$,
so that we can write the approximate equations of motion for the
interstitial vortices as,
\begin{eqnarray}
&&\dot{x}_i +\eta_c^{(0)}\left[1+\frac{1}{2}k^2y_i^2\right]\; \sin\left(x_ik\right) = \eta \label{eq:Eq_11}\\
&&\dot{y}_i +\frac{2\pi}{l_xl_y}y_i = \varepsilon\sin{\Omega t} \; ,
\label{eq:Eq_12}
\end{eqnarray}
where we have retained terms in $y_i$ only to lowest order. We have required
$\pi|y_i|\ll l_x$ in order to obtain
Eqs.~(\ref{eq:Eq_11}) and (\ref{eq:Eq_12}).

The solution to the second of those equations is easily found,
\begin{eqnarray}
y_i & = & \tilde{\varepsilon} \; \sin(\Omega t+\theta)
\label{eq:Eq_13}
\end{eqnarray}
with
\begin{eqnarray}
\tilde{\varepsilon} & = & \frac{\varepsilon}{\Omega\;\sqrt{1+\left(\frac{2\pi}{\Omega l_xl_y}\right)^2}}
\label{eq:Eq_14} \\
\theta & = & \tan^{-1}\frac{2\pi}{\Omega l_xl_y}
\label{eq:Eq_15}
\end{eqnarray}

Inserting Eq.~(\ref{eq:Eq_13}) into Eq.~(\ref{eq:Eq_11}) yields,
\begin{eqnarray}
\dot{x}_i & + & \eta_c^{(0)}\left[1+\frac{1}{4}k^2\tilde{\varepsilon}^2\right]\sin{kx_i} 
\label{eq:Eq_16}\\
& - & \eta_c^{(0)}\frac{1}{4}k^2\tilde{\varepsilon}^2\cos{2(\Omega t+\theta)}\sin{kx_i} = \eta \; .\nonumber
\end{eqnarray}
This equation is the effective equation of motion for the vortex behavior
in the longitudinal ($x$) direction given an ac-force in the transverse ($y$)
direction, derived within the approximations listed above. The equation
is equivalent to that of an overdamped pendulum ($kx_i$ being the pendulum
phase) with dc-torque ($\eta$) and a pivot
vertically oscillating with frequency $2\Omega$ and amplitude proportional to
$\varepsilon^2$ -- i.e., an overdamped
equivalent of the classic Kapitza problem \cite{Kapitza} for underdamped
and parametrically driven pendula.

We will now consider a few separate cases of vortex responses to the
transverse ac-force:

\noindent
{\underline{{\bf $\langle\dot{x}_i\rangle\equiv0$} {$\Leftrightarrow$}
{$x_i(t)=x_0$}:}}
We will omit the last term on the left hand side of Eq.~(\ref{eq:Eq_16}).
The remaining (dc) terms are:
\begin{eqnarray}
\eta_c\sin{kx_0} & = & \eta \label{eq:Eq_17} \\
\Leftrightarrow \; |\eta| & \le & \eta_c = \eta_c^{(0)}\left[1+\frac{1}{4}k^2\tilde{\varepsilon}^2\right] = \eta_c^{(0)}+\delta\eta_c \label{eq:Eq_18} \\
& = & \frac{2\pi}{l_x}{\rm sech}\left(\pi\frac{l_y}{l_x}\right)\left[1+\left(\frac{\varepsilon}{\Omega}\right)^2\frac{\pi^2}{l_x^2}\frac{1}{1+\left(\frac{2\pi}{\Omega l_xl_y}\right)^2}\right] \nonumber \\
\delta\eta_c & = & \eta_c^{(0)}\left(\frac{\varepsilon}{\Omega}\right)^2\frac{\pi^2}{l_x^2}\frac{1}{1+\left(\frac{2\pi}{\Omega l_xl_y}\right)^2} \label{eq:Eq_19}
\end{eqnarray}
where $\eta_c$ is the critical dc force.
It is here worthwhile to notice that the
critical current, $\eta_c$, {\it increases} ($\delta\eta_c$ positive)
quadratically with the transverse ac-amplitude, $\varepsilon$.
This is in direct contrast to the case when the ac-drive is longitudinal,
in which case the critical dc force {\it decreases} quadratically \cite{Barone}.
We remember that validity of the expressions requires
$\pi\varepsilon\ll\Omega{l_x}$.

\noindent
{\underline{{\bf $\langle\dot{x}_i\rangle>0$} {$\Leftrightarrow$}
{$|\eta|>\eta_c$}:}}
Without the parametric term in Eq.~(\ref{eq:Eq_16}), the solution,
for $\eta^2\ge\eta_c^2$, is given by \cite{Barone}
\begin{eqnarray}
v_0 & = & \langle\dot{x}_i\rangle = \sqrt{\eta_0^2-\eta_c^2} \; , \label{eq:Eq_20} \\
\dot{x}_i & = & \eta_0\frac{1-\frac{\eta_c^2}{\eta_0^2}}{1-\frac{\eta_c}{\eta_0}\cos{kv_0t}}\; ,  
\label{eq:Eq_21}
\end{eqnarray}
where $\eta_0$ is the dc-force when the parametric term in Eq.~(\ref{eq:Eq_16})
is neglected.  Let us assume the following simple ansatz for the vortex motion,
\begin{eqnarray}
x_i(t) & = & v_0t+x_0
\label{eq:Eq_22}
\end{eqnarray}
where $x_0$ is a constant. Inserting this ansatz into the effective equation of
motion, Eq.~(\ref{eq:Eq_16}), while only maintaining dc terms,
yields,
\begin{eqnarray}
v_0-\delta\eta_c\langle\cos{2(\Omega t+\theta)}\sin{kx_i(t)}\rangle & = & \eta
= \eta_0+\delta\eta \; .
\label{eq:Eq_23}
\end{eqnarray}
With the ansatz of Eq.~(\ref{eq:Eq_22}) we can therefore expect the term
$\langle\cdot\rangle$ to contribute to $\delta\eta$ if $kv_0=2\Omega$ (see
figure 2), and the resulting relationship between the internal phase,
$kx_0-2\theta$, and the dc current, $\eta_0+\delta\eta$, is:
\begin{eqnarray}
2\Omega/k-\frac{1}{2}\delta\eta_c\sin(kx_0-2\theta) & = & \eta_0+\delta\eta \; .
\label{eq:Eq_24}
\end{eqnarray}
As the phase, $kx_0-2\theta$, can be adjusted to balance this equation
for different choices of $\delta\eta$, we can argue for a nonzero range,
$\Delta\eta$,
in dc-force for which the average speed of the interstitial vortices is
unchanged. This phase-locking range has the magnitude:
\begin{eqnarray}
\Delta\eta & = & \delta\eta_c \; .
\label{eq:Eq_25}
\end{eqnarray}
Thus, we can predict phase-locking in the transversely ac-driven vortex
system and the predicted total range in phase-locking is equal to the increase
in the critical dc-force due to the ac-drive. This prediction is correct
up to and including terms $\propto\varepsilon^2$.

This range in bias current for which the average speed (voltage)
of the interstitial vortex is constant will manifest itself as a step in
the dc current-voltage characteristics of the system. It is important
to emphasize that this step is very different in origin from the Shapiro
steps \cite{Shapiro2} recently demonstrated for longitudinal ac-drive. The
essential difference lies in the fact that the Shapiro steps arise from
a ``direct" driving term in an effective pendulum equation, whereas the
present phase-locking for the transverse ac-drive arises from an effective
{\it parametric} ac-driving term in the longitudinal equation of motion.

If one chooses a better (and more complicated) ansatz for the vortex
motion (e.g., Eq.~(\ref{eq:Eq_21}) instead of Eq.~(\ref{eq:Eq_22})) it
becomes evident that an $\varepsilon^2$ phase-locked step also exists
for $kv_0=\Omega$ and that higher order phase-locked steps may exist at
any sub and super harmonic of the driving frequency. However, we will not
go into detail with other phase-locked modes in this presentation.

\section{Numerical Simulations}

In order to validate the analysis of Sec.III predicting some basic effects of
a transverse ac-force on vortices, we have conducted numerical simulations
of driven interstitial vortices in a rectangular lattice of pinned vortices.

Figure 3 shows a series of simulated (normalized) current (dc-force) voltage
(average vortex velocity, $v_0$) for different values of the transverse
ac-drive. The system is a square array of $4\times4$ pinned vortices with
periodicity $(l_x,l_y)$=$(2,2)$ in a computational simulation box of size
$(L_x,L_y)=4l_x,4l_y)$. The simulated
magnetic field corresponds to $B/B_\phi=1.5$ such that each pinning center is
occupied by a vortex and every other plaquette has an interstitial vortex. The
applied normalized frequency is $\Omega=4$. The four curves (shifted vertically
for clarity) in figure 3 are for
(from top) $\varepsilon=0,1,2,3$. According to the
analysis above, the critical dc-force, $\eta_c$, resulting in onset of voltage
(vortex transport) should increase quadratically
for increasing $\varepsilon$. This is clearly visible from figure 3.
A phase-locked step in dc bias current should develop for
increasing $\varepsilon$ around $kv_0=2\Omega$. This is also clearly
visible for $\eta\approx2.5$. As the ac amplitude is increased we observe
more steps in the IV characteristics. However, while the simplest of these,
$kv_0=\Omega$, can be analyzed in some detail, we will be
concerned with only the steps at $kv_0=0$ and $kv_0=2\Omega$ here.

Two characteristically different vortex evolutions are shown as insets
to figure 3. At bottom right we show a snapshot ($\bullet$) of a vortex
configuration within the parameter range of the phase-locked step for
$\varepsilon=2$ together with a long time trace (thin line) of the vortex
trajectories. The vortex paths are obviously periodic and the interstitial
vortices seem to move in a geometrically simple configuration where all
interactions between the interstitials cancel due to symmetry.
The top left inset shows the traces outside of the phase-locked region.
Here we observe the pinned vortices as static dots, while the interstitials
move in a chaotic or quasiperiodic band between the pinning sites.

A detailed investigation of the validity of the above perturbation results
was performed for several different sets of system parameters. For a system of
a single interstitial vortex, we show in figure 4 the detailed comparisons
between the analytical predictions (lines), Eq.~(\ref{eq:Eq_19}),
and numerical simulations (markers) of the
increase, $\delta\eta_c$, in critical dc force as a function of ac-amplitude.
Figure 4a shows data for square pinning arrays of different plaquette areas
($l_xl_y=4,16$) and with different driving frequencies ($\Omega=4,8$),
while figure 4b shows comparisons for rectangular systems
($l_x/l_y=\frac{1}{2},2$) of one plaquette area ($l_xl_y=4$) for 
different frequencies ($\Omega=4,8$). The details of the parameter sets are
given in the caption. The agreement between simulation and analysis is
obviously overall good for these different parameters. Certainly the
quadratic relationship between the increase of the critical current and
the ac-amplitude is confirmed. The quantitative agreement is also rather
good. The simulation results seem underestimated by the predicted values.
However, this is to be expected since the analysis only includes the transverse
vortex motion as moving in a harmonic potential. The vortex model will
provide a stronger nonlinearity for larger ac-amplitudes, resulting in
a larger than predicted increase of the critical current.

For the same parameter sets as shown in figure 4, we have performed
simulations of the phase-locked step at $kv_0=2\Omega$ and compared the
results to the predicted range, $\Delta\eta$, given in Eq.~(\ref{eq:Eq_25}).
The results, shown in figure 5, are also here in good agreement with the
predictions. Thus, we are confident that the above analysis of the single
interstitial behavior can be a useful tool for understanding phase-locking 
in this system.

Figure 6 shows, again for the same parameter sets as in figure 4, the
comparison between the magnitude of the phase-locked step, $\Delta\eta$,
and the predicted value, but in this case for $B=1.5B_\phi$; i.e., for
one interstitial vortex per two plaquettes of pinned vortices. Once again
we find rather good agreement between simulations and the simple prediction.
However, we notice that rectangular systems with $l_x=2l_y$ (top curves
in figure 6b) cannot produce a measurable phase-locked step for very small
ac-amplitudes.

If one simulates the system for the same parameter values, but for $B=2B_\phi$
(not shown), the agreement will get rather poor for all rectangular cases
where $l_x=2l_y$, and even for some of the square array cases as well. The
reason for these discrepancies is of course found in the assumptions behind
the perturbation analysis. Particularly, the validity of the
assumption of noninteracting interstitial vortices will break down for
densely populated ($B=2B_\phi$) systems since the vortices are mutually
repulsive. The result is sketched in figure 7, where the vortex repulsion
is deforming the geometrically simple lattice into a configuration where
the effective interstitial vortex-vortex interaction is non-zero. The lattice
deformation is characterized by the two parameters, $\xi^{(x)}$ and $\xi^{(y)}$.
Figure 8 shows simulation results for $L_x=4l_x=L_y=4l_y=8$, $\Omega=4$, and
$B=2B_\phi$. The circular markers ($\bullet$) show the simulated range
in phase-locking and the solid straight line represents the prediction
of Eq.~(\ref{eq:Eq_25}). It is apparent that the agreement between simulations
and prediction is only fair ($\Delta\eta\propto\varepsilon^2$) for
relatively large values of $\varepsilon$. Measuring the average (in
time and space) ``phase difference", $\langle\big|\xi^{(x)}\big|\rangle$ and
$\langle\big|\xi^{(y)}\big|\rangle$, we can correlate the magnitude of
lattice deformation with the comparison between simulation and analysis
of phase-locking. The measures of lattice deformation are defined as,
\begin{eqnarray}
\xi^{(x)}_i & = & (x_i\, {\rm mod}\,  l_x) \\
\xi^{(y)}_i & = & (y_i\, {\rm mod}\,  l_y) \\
\Big|\xi^{(x)}_{i,j}\Big| & = & \left\{\begin{array}{ccr}
\Big|\xi^{(x)}_j-\xi^{(x)}_i\Big| & , & \Big|\xi^{(x)}_j-\xi^{(x)}_i\Big| \le \frac{l_{x}}{2} \\
\\
\Big|l_{x}-\Big|\xi^{(x)}_j-\xi^{(x)}_i\Big|\Big| & , & \Big|\xi^{(x)}_j-\xi^{(x)}_i\Big| > \frac{l_{x}}{2} \\
\end{array}\right. \\
\Big|\xi^{(y)}_{i,j}\Big| & = & \left\{\begin{array}{ccr}
\Big|\xi^{(y)}_j-\xi^{(y)}_i\Big| & , & \Big|\xi^{(y)}_j-\xi^{(y)}_i\Big| \le \frac{l_{y}}{2} \\
\\
\Big|l_{y}-\Big|\xi^{(y)}_j-\xi^{(y)}_i\Big|\Big| & , & \Big|\xi^{(y)}_j-\xi^{(y)}_i\Big| > \frac{l_{y}}{2} \\
\end{array}\right. \\
\left\langle\Big|\xi^{(x)}\Big|\right\rangle & = & \left\langle\Big|\xi^{(x)}_{i,j}\Big|\right\rangle_{t,i,j} \\
\left\langle\Big|\xi^{(y)}\Big|\right\rangle & = & \left\langle\Big|\xi^{(y)}_{i,j}\Big|\right\rangle_{t,i,j} \\
\sigma_x & = &\left\langle\sqrt{\left\langle\big|\xi^{(x)}_{i,j}\big|^2\right\rangle_t-\left\langle\big|\xi^{(x)}_{i,j}\big|\right\rangle^2_t}\right\rangle_{i,j} \\
\sigma_y & = &\left\langle\sqrt{\left\langle\big|\xi^{(y)}_{i,j}\big|^2\right\rangle_t-\left\langle\big|\xi^{(y)}_{i,j}\big|\right\rangle^2_t}\right\rangle_{i,j} \\
\end{eqnarray}
Finally, these averages are averaged over the phase-locked step to provide
a single measure for a given $\varepsilon$
($\langle\big|\xi^{(x,y)}\big|\rangle$
and $\sigma_{x,y}$ are typically smallest near the center of the
locking range).
Thus, $0\le\langle\big|\xi^{(x,y)}\big|\rangle\le l_{x,y}/2$ is a measure
of the average spatial lattice deformation and $\sigma_{x,y}$
is a measure of the average temporal fluctuations in the lattice deformation,
$\Big|\xi_{i,j}^{(x,y)}\Big|$.

The minimum averaged
lattice deformation parameters for a given phase-locked step are shown
in figure 8 as solid markers ($\Box$: $\langle\big|\xi^{(x)}\big|\rangle$;
$\triangle$: $\langle\big|\xi^{(y)}\big|\rangle$). We observe large lattice
deformations at small ac-amplitudes, where the agreement between the
observed range of phase-locking is poor, while the deformations ``collapse"
for larger values of $\varepsilon$, thereby giving rise to better agreement
between simulations and analysis. We have also shown the average of
the temporal fluctuations, $\sigma_x$ and $\sigma_y$, in order to
illustrate that the lattice deformations depend on time when
$\langle\big|\xi^{(x)}\big|\rangle$ and $\langle\big|\xi^{(x)}\big|\rangle$
are non-zero. Thus, it indicates that internal modes of the lattice
deformation are excited and may contribute significantly to the
phase-locking range in the low ac-amplitude cases. For large ac-amplitudes
we observe a saturation of the range in phase-locking accompanied by
reappearance of the deformation lattice. This is consistent with chaotic
behavior which are expected for strong ac-driven nonlinear systems.

\section{Conclusion}

We have developed a simple analysis of the behavior of interstitial
vortices in systems with periodic pinning and transverse ac-drive. The
result indicates that the dc-driven longitudinal motion can phase-lock
to the transverse ac-signal and that the range of phase-locking 
(in dc-current) is
quadratic in the ac-amplitude and we have developed a quantitative
expression providing detailed dependencies of also other relevant system
parameters, such as pinning geometry and driving frequency.
We have further demonstrated that the
critical current increases quadratically with the ac-amplitude. The
perturbation results have been validated by numerical simulations which
show good agreement with the analytical predictions in the expected
range of system parameters. The mechanism of phase-locking discussed in
this paper is distinct from phase-locking to a longitudinally applied
ac-signal. The latter case was studied in \cite{Shapiro2} and exhibits
qualitatively different responses to ac-perturbations, such as decreasing
critical dc-current with increasing ac-amplitude and phase-locked steps
that grow linearly with ac-amplitude. The results shown in this paper
are for commensurate fields. Incommensurate fields, where no simple
geometrical relationship can exist between the pinned and interstitial
lattices, are characterized by non-cancelling interactions between
interstitial vortices. Thus, phase-locking ranges for the non-commensurate
fields usually have magnitudes less than predicted by the single
interstitial analysis.

Our results and analyses indicate that the most important phase-locking
appears at relatively small magnetic fields (with respects to the first
matching field) where the inter-vortex repulsion
is not deforming the interstitial vortex lattice.

The predictions in this paper should be directly applicable for experimental
verification in superconductors (or Josephson junction arrays) where
dc and ac fields are orthogonal.

\section{Acknowledgments}

Parts of this work were supported by
ANPCYT (Proy.~03-00000-01034), Fundacion Autochas (Proy.~A-13532/1-96),
Conicet, CNEA and FOMEC (Argentina)
and by the Director, Office of Advanced Scientific Computing Research,
Division of Mathematics, Information and Computing Sciences, U.S.~Department of Energy contract DE-AC03-76SF00098.

\begin{figure}
\caption{Static critical force, $\eta_c^{(0)}\hat{\bf x}$, for $\varepsilon=0$
as a function
of rectangular periodic pinning aspect ratio. Thick lines represent the maximum
force derived from Eq.~(\ref{eq:Eq_8}) for the optimal position, $x_*$,
along the x-axis.
Thin line represent Eq.~(\ref{eq:Eq_9}), which is valid for
${\rm sech}(\pi l_y/l_x)\ll 1$,
dashed line represent $\pi/l_y$, which is asymptotically correct for
$l_x\gg l_y$.
}
\label{fig:fig1}
\end{figure}

\begin{figure}
\caption{Sketch of interstitial vortex ($\bullet$) trajectory in rectangular
lattice of pinned vortices ($\circ$). Upper part of sketch shows a case
where $\langle\dot{x}_i\rangle=0$ while the lower part
shows a case where $\langle\dot{x}_i\rangle=l_x\Omega/\pi$.
Ac-force is transverse, dc-force is longitudinal.
}
\label{fig:fig2}
\end{figure}

\begin{figure}
\caption{Simulated IV characteristics for different values of transverse
ac-amplitude, $\varepsilon=0,1,2,3$ (top down). IV curves are vertically
offset for clarity. System parameters are: $L_x=L_y=4l_x=4l_y=8$,
$\Omega=4$, and $B=1.5B_\phi$. Lower inset shows vortex trajectories
in a phase-locked state while upper inset shows trajectories outside
a phase locked regime.
}
\label{fig:fig3}
\end{figure}

\begin{figure}
\caption{Increase of critical dc current, $\delta\eta_c$ as a function
of ac-amplitude, $\varepsilon$. Markers represent numerical simulations
of equations (\ref{eq:Eq_6}) and (\ref{eq:Eq_7}). Solid lines are the
corresponding predictions from Eq.~(\ref{eq:Eq_19}). Simulations are
conducted for a single interstitial vortex.
(a)
$\triangle$: $l_x=l_y=2$ and $\Omega=4$; 
$\Diamond$: $l_x=l_y=2$ and $\Omega=8$; 
$\Box$: $l_x=l_y=4$ and $\Omega=4$; 
$\circ$: $l_x=l_y=4$ and $\Omega=8$.
(b)
$\triangle$: $l_x=2l_y=4$ and $\Omega=4$; 
$\Diamond$: $l_x=2l_y=4$ and $\Omega=8$; 
$\Box$: $2l_x=l_y=4$ and $\Omega=4$; 
$\circ$: $2l_x=l_y=4$ and $\Omega=8$.
}
\label{fig:fig4}
\end{figure}

\begin{figure}
\caption{Magnitude, $\Delta\eta$, of the phase-locked step at $kv_0=2\Omega$
as a function
of ac-amplitude, $\varepsilon$. Markers represent numerical simulations
of equations (\ref{eq:Eq_6}) and (\ref{eq:Eq_7}). Solid lines are the
corresponding predictions from Eq.~(\ref{eq:Eq_25}). Simulations are
conducted for a single interstitial vortex.
(a)
$\triangle$: $l_x=l_y=2$ and $\Omega=4$; 
$\Diamond$: $l_x=l_y=2$ and $\Omega=8$; 
$\Box$: $l_x=l_y=4$ and $\Omega=4$; 
$\circ$: $l_x=l_y=4$ and $\Omega=8$.
(b)
$\triangle$: $l_x=2l_y=4$ and $\Omega=4$; 
$\Diamond$: $l_x=2l_y=4$ and $\Omega=8$; 
$\Box$: $2l_x=l_y=4$ and $\Omega=4$; 
$\circ$: $2l_x=l_y=4$ and $\Omega=8$.
}
\label{fig:fig5}
\end{figure}

\begin{figure}
\caption{Magnitude, $\Delta\eta$, of the phase-locked step at $kv_0=2\Omega$
as a function
of ac-amplitude, $\varepsilon$. Markers represent numerical simulations
of equations (\ref{eq:Eq_6}) and (\ref{eq:Eq_7}). Solid lines are the
corresponding predictions from Eq.~(\ref{eq:Eq_25}). Simulations are
conducted for $B=1.5B_\phi$, $L_x=4l_x$, and $L_y=4l_y$.
(a)
$\triangle$: $l_x=l_y=2$ and $\Omega=4$; 
$\Diamond$: $l_x=l_y=2$ and $\Omega=8$; 
$\Box$: $l_x=l_y=4$ and $\Omega=4$; 
$\circ$: $l_x=l_y=4$ and $\Omega=8$.
(b)
$\triangle$: $l_x=2l_y=4$ and $\Omega=4$; 
$\Diamond$: $l_x=2l_y=4$ and $\Omega=8$; 
$\Box$: $2l_x=l_y=4$ and $\Omega=4$; 
$\circ$: $2l_x=l_y=4$ and $\Omega=8$.
}
\label{fig:fig6}
\end{figure}

\begin{figure}
\caption{Sketch of interstitial vortex lattice deformed due to the
repulsive vortex-vortex interaction at $B=2B_\phi$. The parameters, $\xi^{(x)}$
and $\xi^{(y)}$, are measures of the magnitude of deformation from a rectangular
lattice.
}
\label{fig:fig7}
\end{figure}

\begin{figure}
\caption{Magnitude, $\Delta\eta$, of the phase-locked step at $kv_0=2\Omega$
as a function
of ac-amplitude, $\varepsilon$, for $L_x=4l_x=L_y=4l_y=8$, $\Omega=4$,
and $b=2B_\phi$ ($\bullet$). Markers represent numerical simulations
of equations (\ref{eq:Eq_6}) and (\ref{eq:Eq_7}). Solid line represent the
corresponding prediction from Eq.~(\ref{eq:Eq_25}). Also shown are the
lattice deformation measures, $\langle\big|\xi^{(x)}\big|\rangle$ and
$\langle\big|\xi^{(y)}\big|\rangle$, as well as their standard deviations,
$\sigma_x$ and $\sigma_y$.
}
\label{fig:fig8}
\end{figure}

\begin{references}
\bibitem{Shapiro} S.~Shapiro, 
Phys.~Rev.~Lett.~{\bf 11}, 80 (1963).

\bibitem{Barone}
A.~Barone and G.~Paterno, {\it Physics and Applications of the Josephson Effect}
(Wiley, New York, 1982).

\bibitem{ShapiroN}
S.P.~Benz, M. S. Rzchowski, M. Tinkham, and C. J. Lobb, 
Phys.~Rev.~Lett.~{\bf 64}, 693 (1990);
K.\ H.\ Lee, D.\ Stroud, and J.\ S.\ Chung, Phys.\ Rev.\
Lett.\ {\bf64}, 692 (1990); K.\ H.\ Lee and D.\ Stroud, Phys.\ Rev.\ B 
{\bf 43},5280 (1991); J.\ U.\ Free, S.\ P.\ Benz, M.\ S.\ Rzchowski,
M.\ Tinkham, C.\ J.\ Lobb, and M.\ Octavio, Phys.\ Rev.\ B {\bf 41},7267
(1990); M.\ Octavio, J.\ U.\ Free, S.\ P.\ Benz, R.\ S.\ Newrock, D.\ B.\
Mast, and C.\ J.\ Lobb, {\it ibid.} {\bf 44},4601 (1991);
D.~Dom{\' \i}nguez and J.V.~Jose, Phys.~Rev.~Lett.~{\bf 69}, 514 (1992).

\bibitem{VanLook}
L.~Van Look, E. Rosseel, M. J. Van Bael,
 K. Temst, V. V. Moshchalkov, and Y. Bruynseraede,
  Phys. Rev. B {\bf 60}, R6998 (1999). 
  
\bibitem{Shapiro2}
C.~Reichhardt, R.T.~Scalettar, G.T.~Zim\'anyi, and N.~Gr{\o}nbech-Jensen,
Phys.~Rev.~B R11 914 (2000).

\bibitem{Daldini}
P.~Martinoli, O.~Daldina, C.~Leemann, and E.~Stocker,
Solid State Commun.~{\bf 17}, 205 (1975).

\bibitem{CDW} G. Gr\"uner, Rev.~Mod.~Phys {\bf 60}, 1129 (1988);
S. Bhattacharya, M.H.~Higgins, J. P. Stokes, and R. A. Klemm, Phys.~Rev.~Lett.
{\bf 59}, 1849 (1987); M.H.~Higgins, A. Alan Middleton, and S. Bhattacharya, 
{\it ibid.} {\bf 70}, 3784 (1993);
S. N. Coppersmith and
P. B. Littlewood, {\it ibid.} {\bf 57}, 1927 (1986);
A. A. Middleton, O. Biham, P. Sibani, and P. B. Littlewood, {\it ibid.} {\bf 68}, 1586 (1992). 

\bibitem{Fiory} A.T.~Fiory, Phys.\ Rev.\ Lett.\ {\bf 27},
501 (1971); J. M. Harris, N. P. Ong, R. Gagnon, and L. Taillefer, 
Phys. Rev. Lett. {\bf 74}, 3684 (1995).
 
 
\bibitem{Kolton} A. B. Kolton, D. Dom\'{\i}nguez, and N.Gr{\o}nbech-Jensen,
Phys. Rev. Lett. {\bf 86}, 4112 (2001).

\bibitem{Metlushko}
M. Baert, V. V. Metlushko, R. Jonckheere, V. V. Moshchalkov,
and Y. Bruynseraede, Phys. Rev. Lett {\bf 74}, 3269 (1995); 
A. Castellanos, R. Wondenweber, G. Ockenfuss, A. v.d. Hart, 
and K. Keck, Appl. Phys. Lett. {\bf 71}, 962 (1997);
V. Metlushko, U. Welp, G. W. Crabtree, Zhao Zhang, S. R. J. Brueck,
B. Watkins, L. E. DeLong, B. Ilic, K. Chung,
 and P. J. Hesketh, Phys. Rev. B {\bf 59}, 603 (1999);
S. B. Field, S. S. James, J. Barentine, 
V. Metlushko, G. Crabtree, H.Shtrikman, B. Ilic, S. R. J. Brueck
, cond-mat/0003415; A.N.~Grigorenko {\it et al}, Phys.~Rev.~B {\bf 63},
52504 (2001).


\bibitem{Moshchalkov}
E. Rosseel, M. Van Bael, M. Baert, R. Jonckheere,
 V. V. Moshchalkov, and Y. Bruynseraede, Phys. Rev. B {\bf 53}, R2983, (1996);
 
\bibitem{Harada}
K. Harada, O. Kamimura, H. Kasai, T. Matsuda, A. Tonomura
and  V. V. Moshchalkov, Science {\bf 274}, 1167 (1996);

\bibitem{Schuller}
J. I. Mart\'{\i}n, M. V\'elez, J. Nogu\'es, and Ivan K.
Schuller , Phys. Rev. Lett. {\bf 79}, 1929 (1997);
D.J.~Morgan and J.B.~Ketterson, Phys.~Rev.~Lett.~{\bf 80}, 3614 (1998);
A.~Hoffmann, P.~Prieto and I.K.~Schuller, Phys.~Rev.~B {\bf 61}, 6985 (2000).;
A.~Terentiev {\it et al.} Physica C {\bf 324}, 1 (1999).

\bibitem{Hoffman}
J.I.Mart\'{\i}n, M. V\'elez, A. Hoffmann, Ivan K. Schuller,
 and J. L. Vicent, Phys. Rev. Lett. {\bf 83}, 1022 (1999);
J.I.~Mart\'{\i}n, M. V\'elez, A. Hoffmann, Ivan K. Schuller, J. L. Vicent, 
Phys.~Rev.~B {\bf 62}, 9110 (2000); S.~Kolesnik {\it et al.},
Physica C {\bf 341-348}, 1093 (2000).

\bibitem{DrivenShort}
C.~Reichhardt, C.J.~Olson, and F.~Nori, Phys. Rev. Lett. {\bf 78}, 2648 
(1997). 

\bibitem{Marconi}
V.I.~Marconi and D.~Dom\'{\i}nguez, Phys.~Rev.~Lett.~{\bf 82}, 4922 (1999).

\bibitem{Jensen2}
N.~Gr{\o}nbech-Jensen, Int.~J.~Mod.~Phys.~C {\bf 7}, 873 (1996);
Comp.~Phys.~Comm.~{\bf 119}, 115 (1999).

\bibitem{Kapitza}
see, e.g., Landau \& Lifshitz, {\it Mechanics} p.~93, 3rd ed.,
Pergamon Press, 1976; J.\ A.\ Blackburn, H.\ J.\ T.\ Smith, and
N.\ Gr{\o}nbech-Jensen, Am.~J.~Phys.~{\bf 61}, 475 (1993).
\end{references}
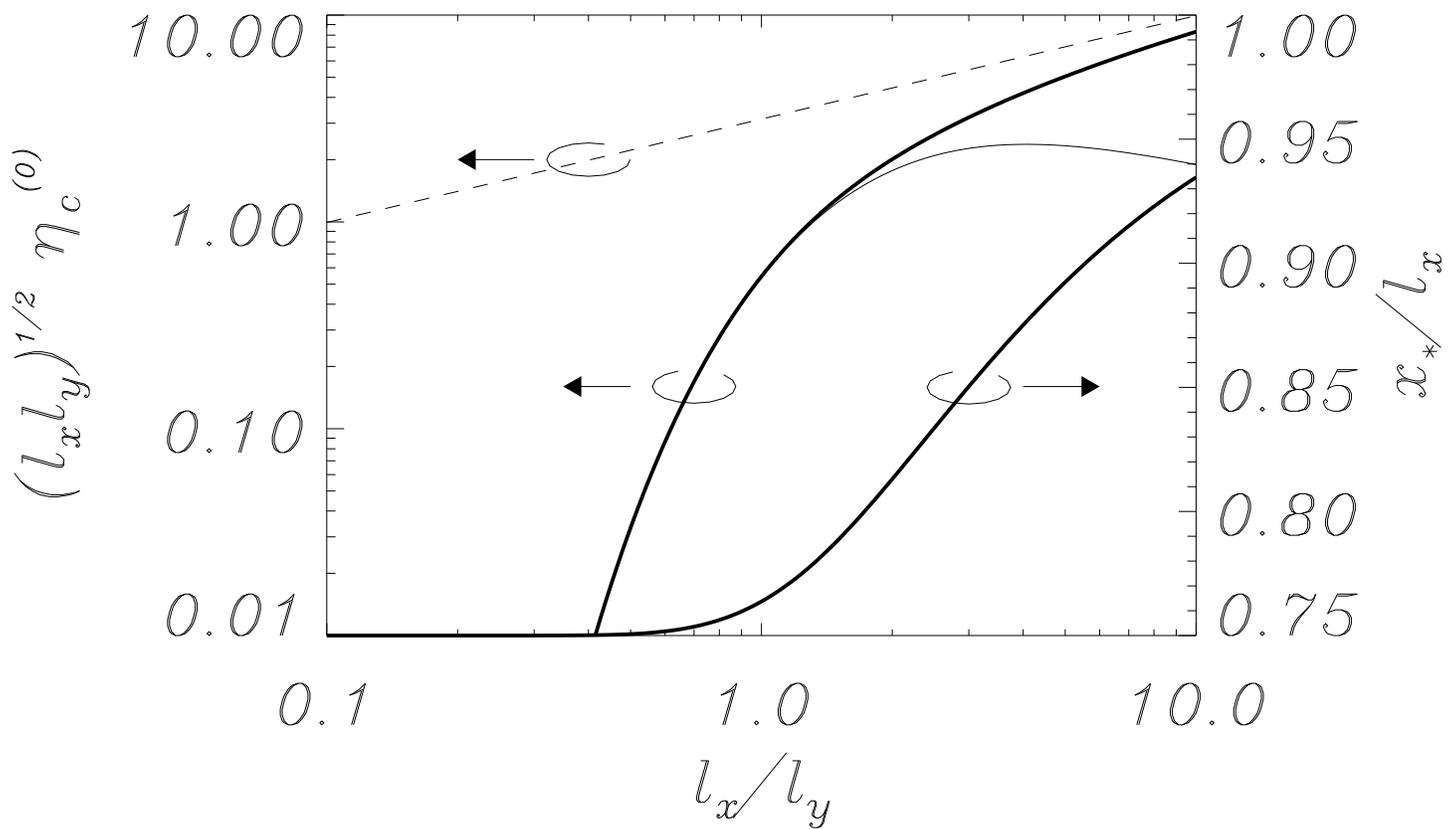
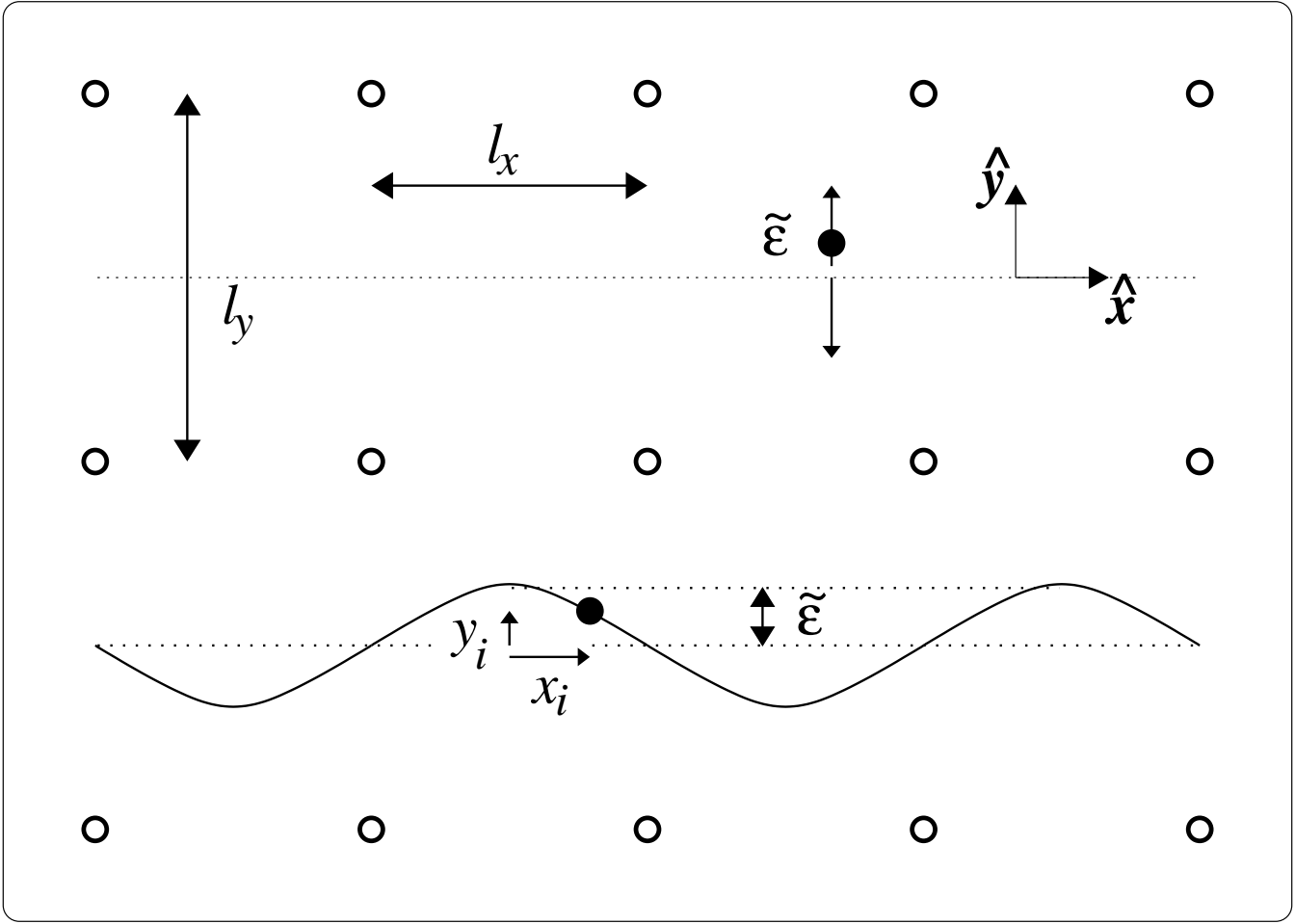
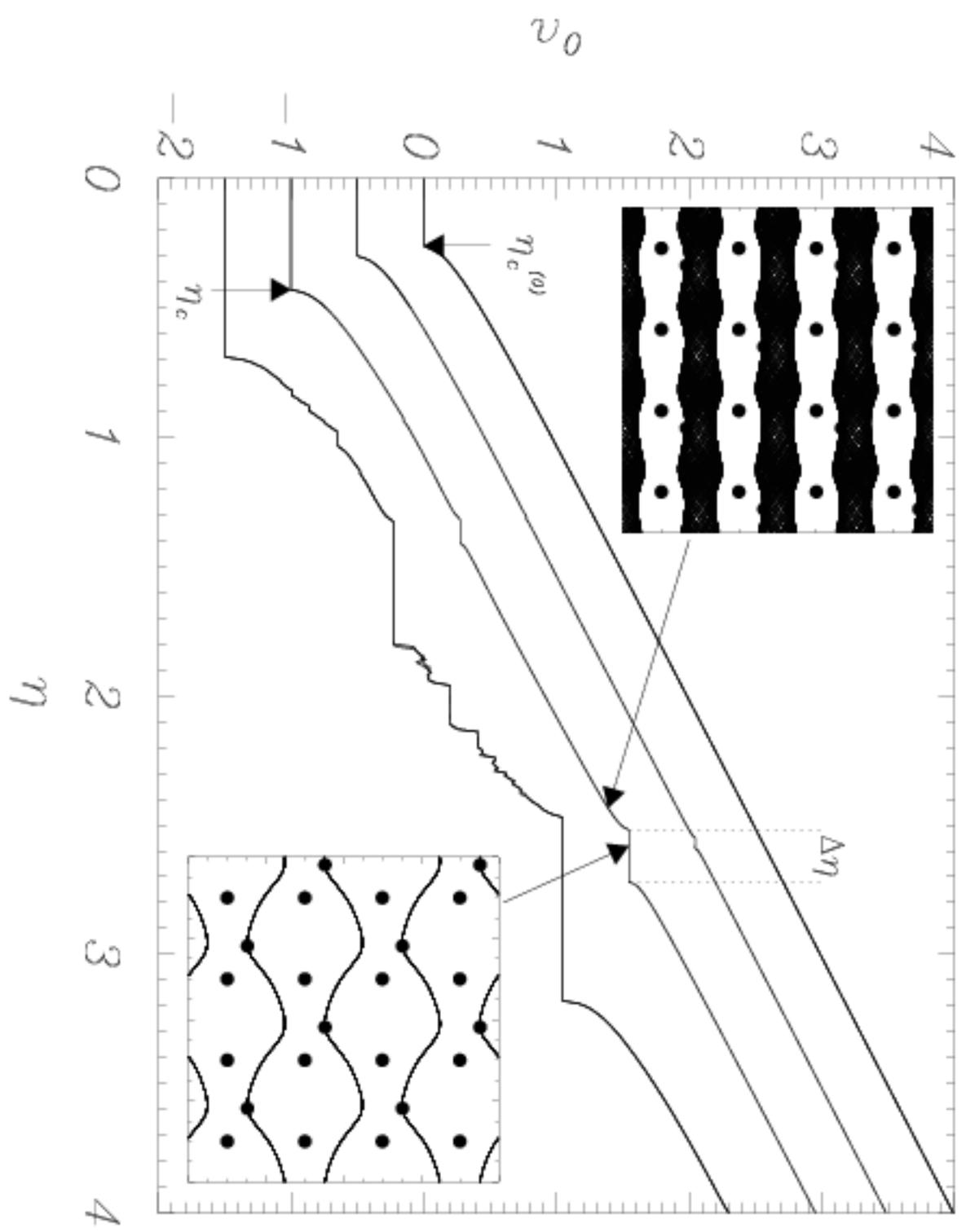
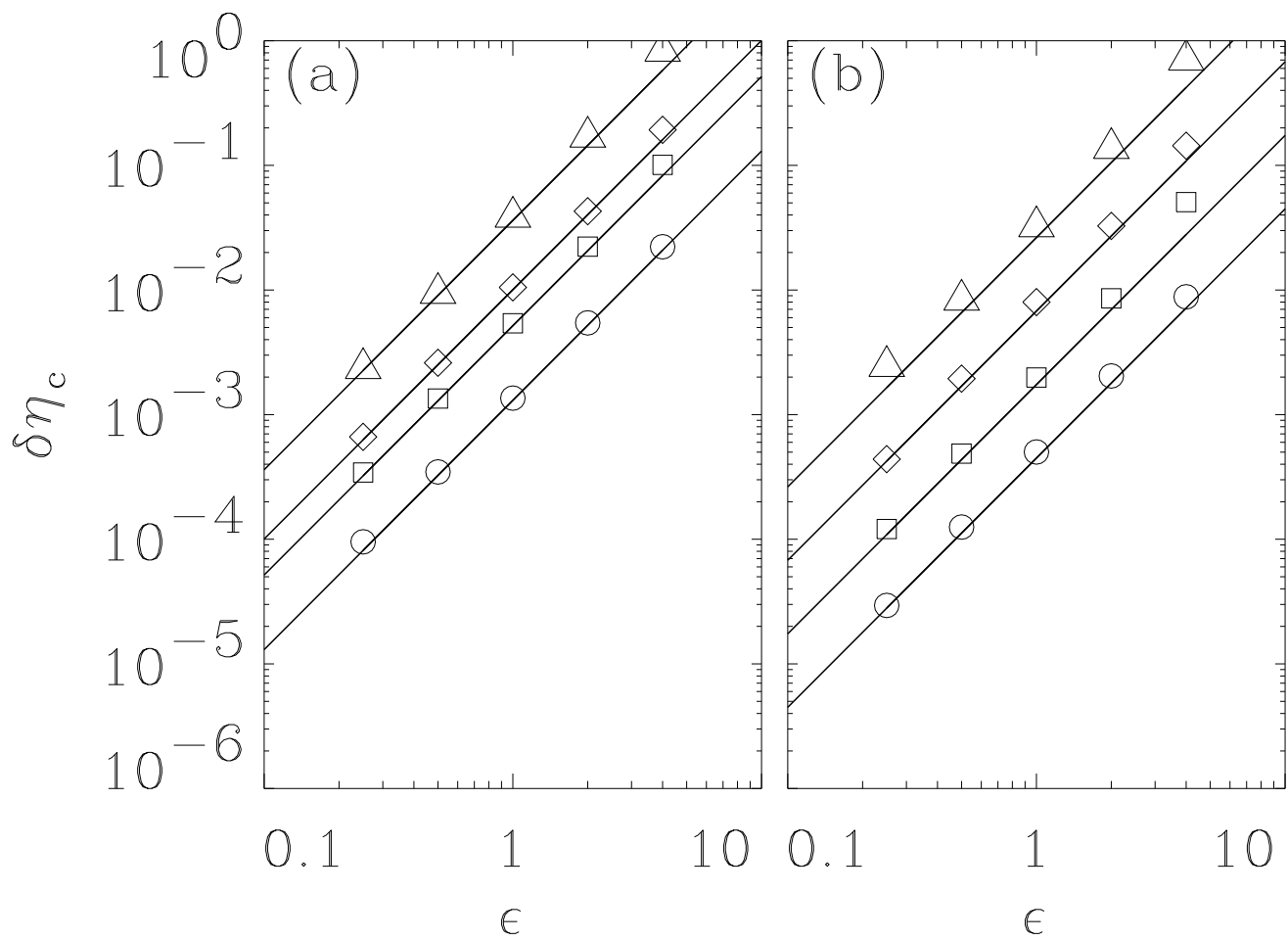
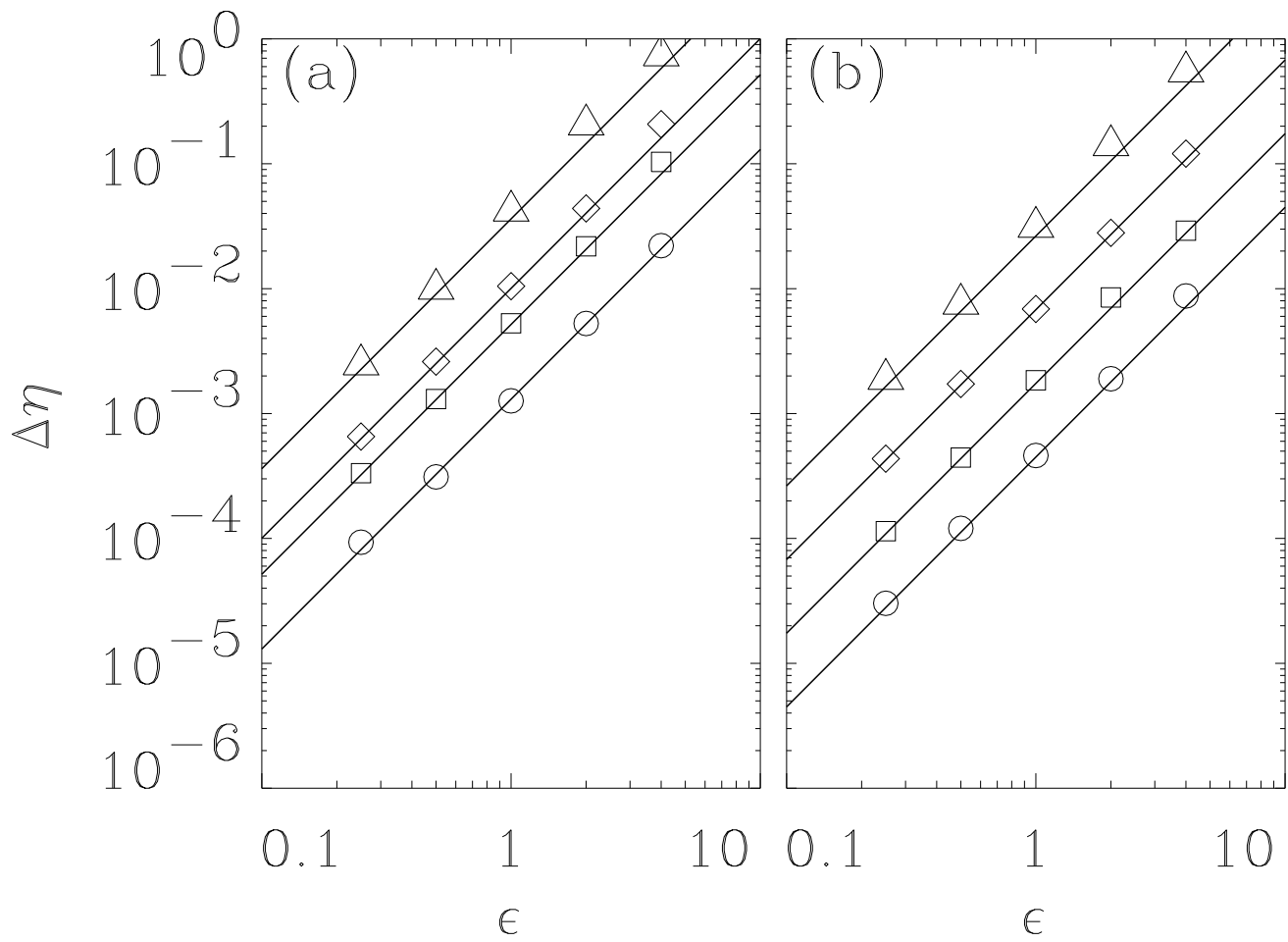
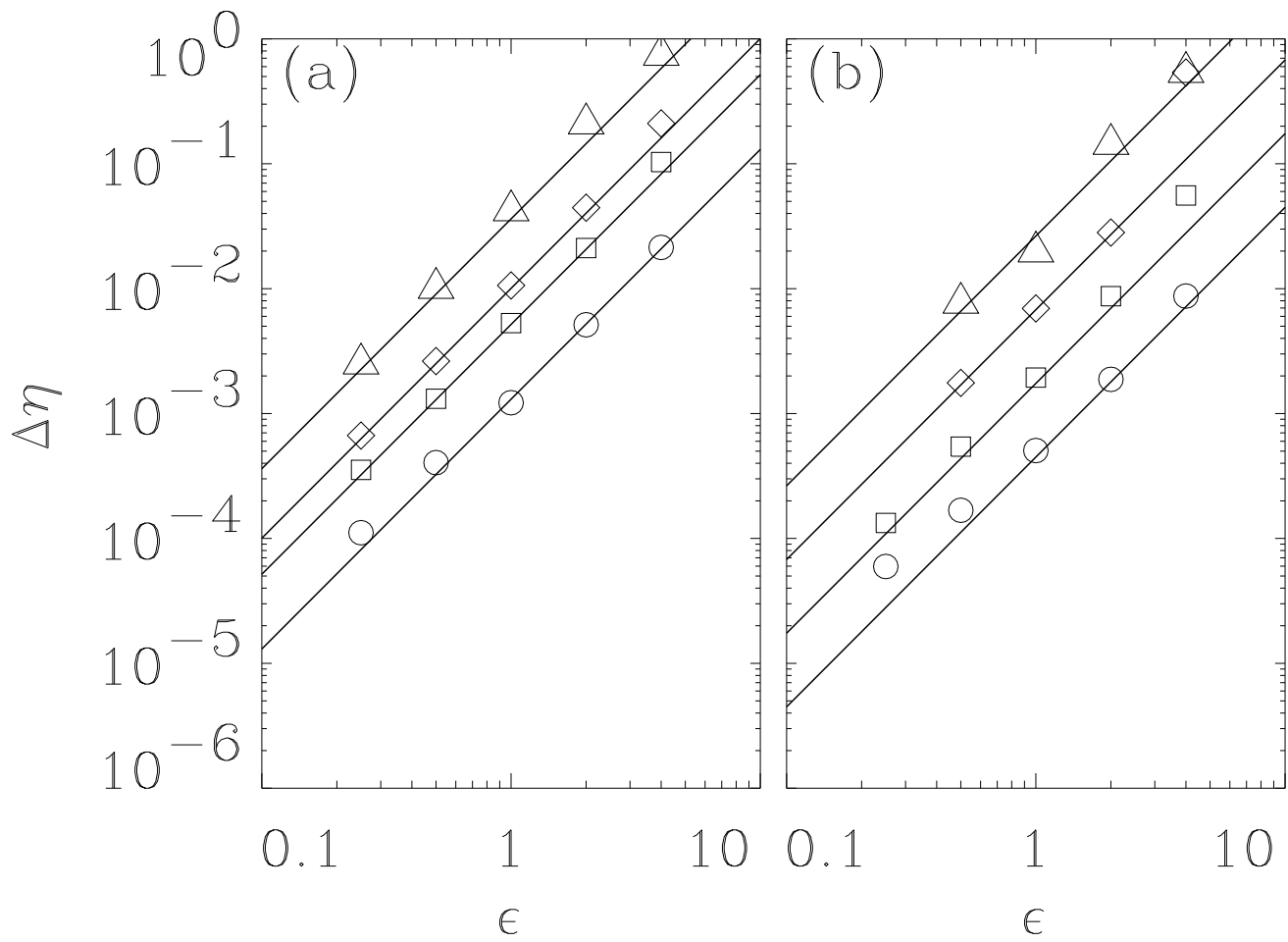
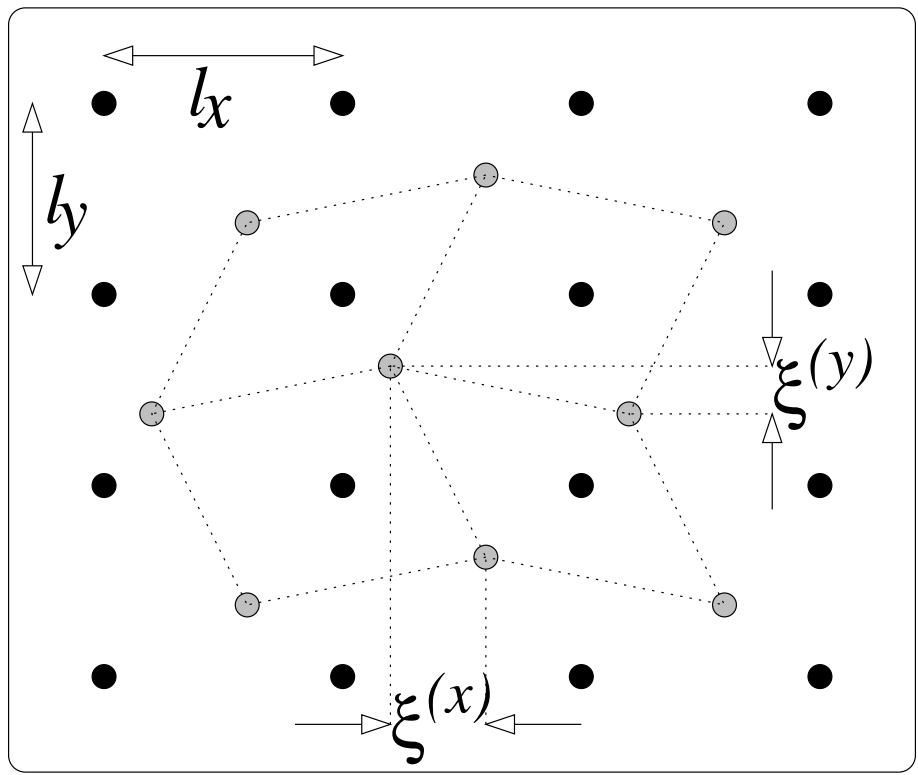
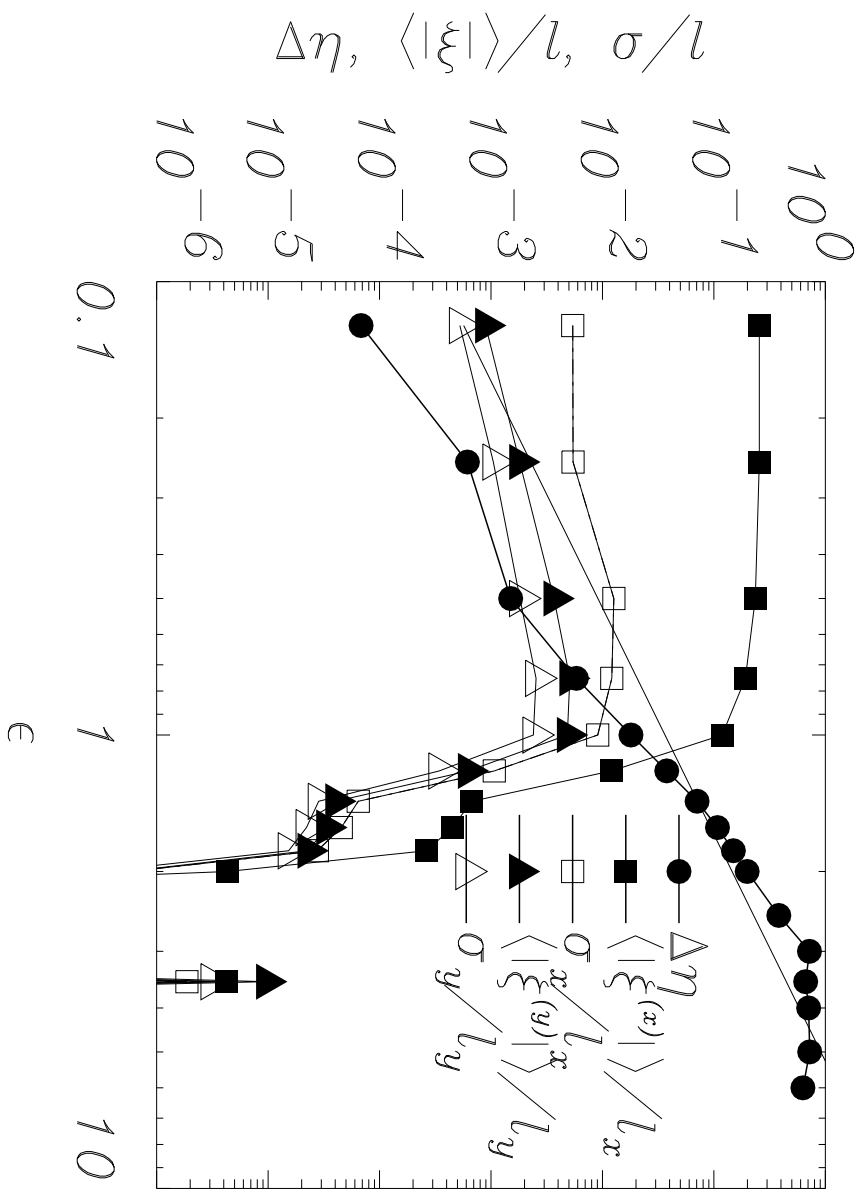
\end{document}